\documentclass[
    prd,
    reprint,
    amsmath,
    amssymb,
    floats,
    superscriptaddress,
    nofootinbib,
    showpacs
]{revtex4-2}

\usepackage{multirow}
\usepackage{graphicx}
\usepackage{subfigure}
\usepackage{sidecap}
\usepackage[normalem]{ulem}
\usepackage{diagbox}
\usepackage{xcolor}

\usepackage{subcaption}
\DeclareCaptionJustification{justified}{\justifying}
\usepackage{hyperref} 

\begin{document}
\author{Daniel~Lanchares}
\email{(first author) daniellanchares@gmail.com}
\affiliation{Departamento de Fisica, Universidad de Oviedo, C. Federico Garcia Lorca 18, 33007 Oviedo, Spain}
\affiliation{Instituto Universitario de Ciencias y Tecnologías Espaciales de Asturias (ICTEA), C. Independencia 13, 33004 Oviedo, Spain}

\author{Lysiane Mornas}
\email{mornas@uniovi.es}
\affiliation{Departamento de Fisica, Universidad de Oviedo, C. Federico Garcia Lorca 18, 33007 Oviedo, Spain}
\affiliation{Instituto Universitario de Ciencias y Tecnologías Espaciales de Asturias (ICTEA), C. Independencia 13, 33004 Oviedo, Spain}

  \author{Luigi Toffolatti}
\email{ltoffolatti@uniovi.es}
\affiliation{Departamento de Fisica, Universidad de Oviedo, C. Federico Garcia Lorca 18, 33007 Oviedo, Spain}
\affiliation{Instituto Universitario de Ciencias y Tecnologías Espaciales de Asturias (ICTEA), C. Independencia 13, 33004 Oviedo, Spain}

\author{Pietro Vischia}
\email{(corresponding author) pietro.vischia@cern.ch}
\affiliation{Departamento de Fisica, Universidad de Oviedo, C. Federico Garcia Lorca 18, 33007 Oviedo, Spain}
\affiliation{Instituto Universitario de Ciencias y Tecnologías Espaciales de Asturias (ICTEA), C. Independencia 13, 33004 Oviedo, Spain}  

\title{Co-design of ground-based gravitational wave detector networks}

\begin{abstract}
Discussions around the design philosophy and location of the next generation of ground-based gravitational wave detectors are still underway. In this context, we propose IfoScout, an innovative methodology for detector co-design based on state-of-the-art machine-learning (ML) techniques. As a simple demonstration, we present a two-stage simulation of a network of fictional L-shaped interferometers whose sensitivity is optimized within physical and geographical constraints, indirectly resulting in reducing the costs. To achieve this, we gather publicly available data for two token locations and establish the length and orientation with reinforcement learning (RL). Next, we optimize the internal detector parameters related to cavity stability to achieve the best possible sensitivity by means of differential programming (DP). We make the case that pipelines such as IfoScout could have a positive impact on the final design of new generation detectors (\textit{e.g.} the Einstein Telescope, the Cosmic Explorer, etc.), given precise data and detailed, realistic simulations of the interferometers.
\end{abstract}

\maketitle

\section{Introduction}
\label{sec:intro}

A decade of gravitational wave (GW) observations \cite{GW150914, GW170817, GWTC1_paper, GWTC-2.1, GWTC-3, GWTC-4.0, GWTC-5.0} in which Advanced LIGO~\cite{Advance_LIGO_paper}, Advanced Virgo~\cite{Advance_Virgo_paper} and more recently KAGRA~\cite{KAGRA_paper} have resulted in valuable discoveries~\cite{basics-of-GW150914, GW170814, GW170817_properties, LVK+Fermi+IceCube, GW231123, GW241011_&_GW241110, GW250114}, has generated a strong case for new, more powerful detectors. Multiple proposals for ground-based detectors such as the Einstein Telescope (ET~\cite{ET}), Cosmic Explorer (CE~\cite{CE_main}) or, more recently, lunar-based detectors (LGWA~\cite{LGWA_2021}, LILA~\cite{LILA_2025}, CIGO~\cite{Zhang2026_CIGO}) are in various stages of development, alongside space-based observatories (LISA~\cite{LISA_2017}, TianQin~\cite{Luo_2016_tianqin}, DECIGO~\cite{Shuichi-Sato_2009_DECIGO}). 

These instruments are set to be some of the most precise ever built, but this precision often comes at the expense of ever-increasing design complexity, making their undertaking a superhuman task, even for large collaborations. Optimization is one of the greatest hurdles of the design process, given the large dimensionality of the space of possible choices for geometry, detection technology, materials, data-acquisition, and information-extraction techniques, and the interdependence of the related parameters.
Many efforts have been carried out to improve specific scientific goals (\textit{e.g.} better sky localization, lower noise at low or high frequencies)~\cite{Miao_2014_small_detector_optim, PhysRevD.99.102004_Detector_optim_for_BNS,Iacovelli2026_baseline_sky_position, Science_ET_2026, Santoliquido_2026_Next_gen_NPE}. 
There have also been many studies on detector site selection~\cite{ET_site_criteria, Burchartz2025_ET_siting_&_design_Rhine, Bristol_CE_site_analysis_2026, Datrier_2026_CE_sites} and how to mitigate its impact on sensitivity~\cite{amann_2022_tunnel_optim, hild2011sensitivity, driggers_2012_newtonian_noise, Harms2019_terrestrial_review, schillings2026numerical}
However, the two groups of studies are mostly separate from each other. A handful of site-independent configurations (sometimes only one) are decided upon first based mainly on scientific goals, and studies on their integration into potential sites are carried out afterwards.

A new paradigm has been recently proposed~\cite{MODE_white_paper}, where the scientific goals of the instrument (\textit{e.g.} quality of the statistical inference on the parameters of interest) and the challenges that constrain its realization (physical and geometrical boundaries, construction costs, etc.) are both encoded into a utility function whose supremum can be found with computational methods. This constrained optimization would allow a systematic search in the detector configuration space, likely achieving sizeable gains in performance at a fraction of the cost. When the simultaneous optimization relates hardware (detector-related) and software (inference-related) parameters, we refer to it as \textit{codesign}~\cite{MODE:2026xoy}. The construction of a fully differentiable pipeline and the use of deep learning techniques could then enable simultaneous optimization of all -continuous- design parameters. Such approaches belong to the framework of \textit{differential programming} (DP), with pre-existing efforts in the GW community focusing around interferometry~\cite{Krenn_digital_discovery, Ruiz-Gonzalez_2026_detector_surrogates_Krenn} and waveform modeling~\cite{PhysRevD.110.064028_DP_waveforms}.

However, meeting such differentiability conditions is hardly possible in practical applications, and GW detectors are no different: their location (and therefore sensitivity) is often constrained by geographical, economical (here represented by underground and supported fractions of the arms), and political limitations that are impractical to model by functions with computable gradients. Besides, detector design requires discrete choices of materials, configuration, etc, which are also non-differentiable.

This hurdle can be overcome by employing reward-based machine learning (ML) paradigms that enable models to learn from non-differentiable simulations of the detector. In exchange for this flexibility, learning speed significantly degrades, as some degree of random exploration of the parameter space will be necessary for meaningful learning, and the two simulations have to be sequential. Nevertheless, methodologies such as these, which belong to the realm of \textit{reinforcement learning} (RL), have proven to be of great use to the community for daily detector operation~\cite{Buchli_2025_RL_for_LIGO_control} and can provide an effective fallback for cases where full differentiability of the pipeline is not feasible, as illustrated by the hybrid differentiable/RL optimization of the Southern Wide-field Gamma-ray Observatory~\cite{MODE:2026xoy}.

\begin{figure}[b]
      \centering
      \includegraphics[width=0.5\textwidth]{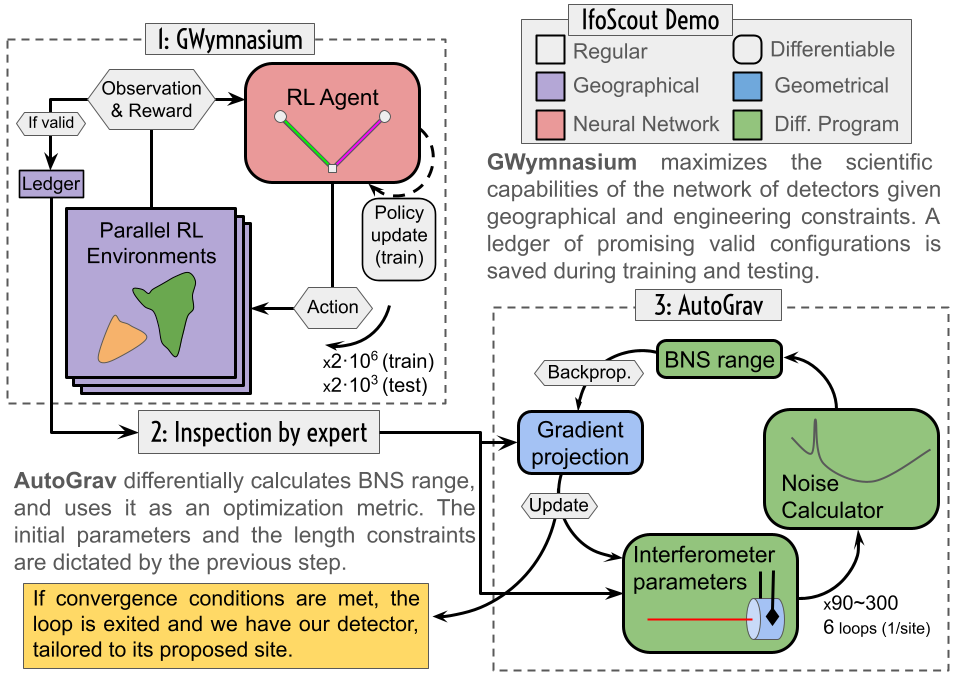}
      \captionof{figure}{Diagram of the \texttt{IfoScout} pipeline detailing the optimization loops of the two sequential stages.}
    \label{fig:diagram}
\end{figure}

In this work, we introduce \texttt{IfoScout} (\textit{Interferometer Scout}, Fig.~\ref{fig:diagram}), a novel approach to GW detector design that could meaningfully impact the development phase of third generation ground-based detectors. By combining differentiable programming~\cite{blondel-2025-elements-differentiable-programming} with reinforcement learning~\cite{RL_atari} on a multi-stage optimization, we are able to arrive at detector configurations that optimize scientific goals while respecting human-imposed boundaries. This pipeline fits neatly into the general framework recently laid out in~\cite{Klimesch2026_Nature_review}. 

To apply differential programming to GW detectors within \texttt{IfoScout}, we present \texttt{AutoGrav}, a translation to \texttt{PyTorch}~\cite{pytorch_paper} of two common GW libraries, \texttt{PyGWINC} and \texttt{inspiral\_range}~\cite{ligogwinc}, that enables gradient computation of their main routines. 
Leveraging pre-existing standards in reinforcement learning \cite{stable-baselines3-RL-2021, gymnasium-RL-2025} we designed flexible environments that simulate the interferometer with a macro level of detail. We refer to this portion of the pipeline as \texttt{GWymnasium}.

In this proof-of-concept paper, we do not pretend to handle the full problem, but to demonstrate the potential benefits of such a method. To illustrate this, we study possible networks of two L-shaped detectors in two plausible sites in the Spanish countryside based on real topographic surveys. After running \texttt{GWymnasium}, the best (\textit{i.e.} most rewarded) configurations are manually inspected and filtered, and those selected are optimized in \texttt{AutoGrav}.
Even in this simplified treatment, we find three network configurations that exploit the geographic features of the sites to produce large detectors with a relatively low impact on civil infrastructure.
Although in this paper we limit ourselves to a network of two detectors, \texttt{IfoScout} natively supports an arbitrary number of detectors, limited only by computational considerations.

\section{Methodology}
\label{sec:methodology}

\subsection{Differential programming: AutoGrav}\label{subsec:dp}

When discussing potential future GW observatories, sensitivity curves are an essential component of any proposal~\cite{ET, LISA_2017}. The curves represent, for a given frequency, the contribution of individual noise sources to the amplitude spectral density of the data that will be obtained. They are a metric that determines which sources are potentially detectable and quantifies the scientific impact of any engineering decision as a function of detector parameters. Therefore, they are a prime utility function candidate for detector optimization~\cite{MODE_white_paper}. 

These curves are computed, as a first approximation, with the \texttt{PyGWINC}~library~\cite{ligogwinc}.  A figure of merit can be calculated from the combined contribution of all noise sources, the SenseMonitor range~\cite{sensemon_range}. It is an approximate measure of the maximum distance at which binary neutron star coalescences (BNSs) can be confidently detected (signal-to-noise ratio larger than 8~\cite{Fairhurst_2011_justif_snr_8_bns_range}). Although more precise range definitions exist, we are only interested in maximization, not in a specific range. 

To use sensitivity curves in optimization, we have created \texttt{AutoGrav}. Its core functionality emulates \texttt{PyGWINC} routines within the \texttt{torch} ecosystem. With it, we are able to load an initial configuration as a starting point (in our case, Adv. LIGO~\cite{adv_ligo_design_yaml}), compute the gradient of the BNS range with respect to every parameter of interest, and iteratively perform gradient descent. Since some parameters are constrained by physical boundaries (\textit{e.g.} the optical stability of the cavities), we also implement protocols to tackle constrained optimization problems~\cite{1965_complex_constrained_optimization,MODE:2026xoy}, modifying the standard gradient descent algorithm in such a way that it can properly explore regions of the phase space arbitrarily close to physical boundaries. We achieve this by smoothly projecting out gradient components non-perpendicular to a planar boundary. We can remain arbitrarily close to it, as outlined in~Eq.~\ref{eq:grad_projection}, where $f$ if the loss function, $\vec{n}$ the normal to the plane, $g$ is the g-factor defined in ~Eq. \ref{eq:g-factor} and the projection starts when $g_{ini}$ is reached and forces convergence to $g_{tgt}$.
\begin{equation}\label{eq:grad_projection}
    \vec{\nabla}f_{\parallel} = \vec{\nabla}f - \alpha(g)(\vec{\nabla}f\cdot\vec{n})\vec{n} \hspace{0.3cm} \begin{cases}
        \alpha(g_{ini}) = 0\\
\alpha(g_{tgt}) = 1
    \end{cases}
\end{equation}
\begin{equation}\label{eq:g-factor}
     g^{2} := g_1g_2(1-g_1g_2) \geq 0,
     \begin{cases}
     g_1 := 1-L/R_{ITM}\\ g_2 := 1-L/R_{ETM}\;, 
\end{cases}
\end{equation}
This strategy allows the teams to agree on a target $g_{tgt}$ prior to optimization. For this simple demo, the main parameters are the radius of the mirrors of the Input and End Test Masses (ITM, ETM). For better convergence, arm length is also varied up to the limit set by GWymnasium (for details, see Supplemental Material).

\begin{figure*}[t]
      \centering
      \includegraphics[width=\textwidth]{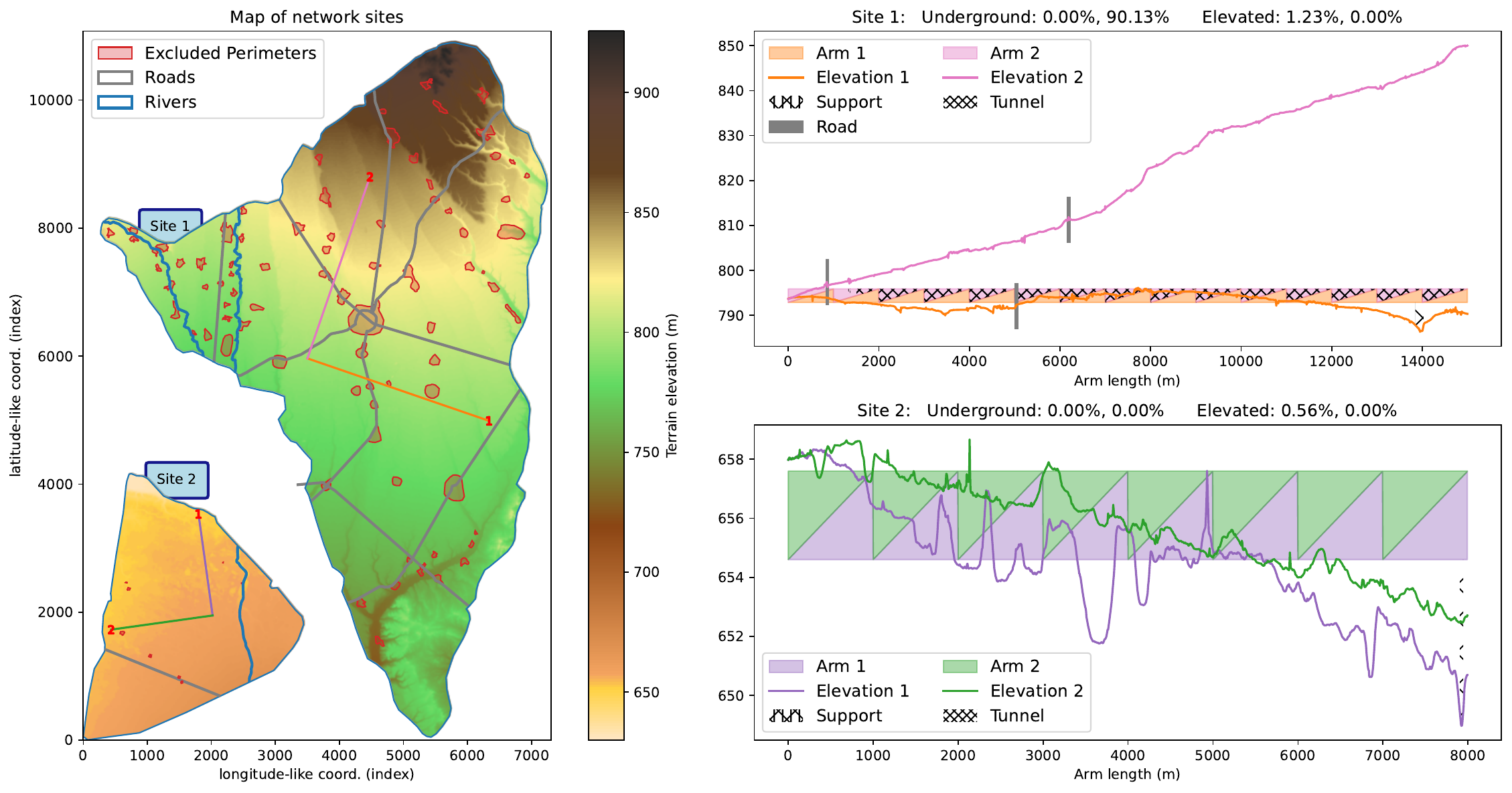}
      \captionof{figure}{\textbf{Example configuration of the network of gravitational wave detectors.} \textbf{Left:} Map of the two chosen sites (site 1: Páramo Leonés, site 2: Alcázar de San Juan) with their respective interferometers in Configuration 2 (see Results). Red shaded regions are areas to avoid. \textbf{Right:} Elevation profiles of the terrain in the paths of the interferometer  arms (up: site 1, down: site 2) with underground/elevated portions marked by hatching (this configuration requires almost not bridgework. Hatching is more visible in Fig.~\ref{fig:configuration_3}). A buffer of 5 meters above and below the arms (whose height is set to 3 meters) is considered before resorting to tunnels or supports. Vertical lines represent intersections between roads and the terrain, and their width has been exaggerated for clarity.}
    \label{fig:configuration}
\end{figure*}

\subsection{Reinforcement learning: GWymnasium}  \label{subsec:rl}
If left unconstrained, \texttt{AutoGrav} will (trivially, in this simplified setting\footnote{Supposing there are no constraints on the maximum length such as those introduced \textit{e.g.} in Refs.~\cite{PhysRevD.99.102004_Detector_optim_for_BNS,Ackley:2020atn}: we address a possible extension in the discussion}) infer that the most direct route to improving detector sensitivity is lengthening the detector arms (larger $L$). While this is an essential part of third generation detector designs compared to their predecessors, detectors cannot be enlarged \textit{ad-infinitum} in a finite Earth with finite resources.
To address this, we reformulate the process of finding the best spatial configuration of a network of interferometers by introducing a number of constraints in the RL loop.
The real-world environment of the detectors can naturally be simulated on a computer through the use of geographical (and, if available, geological) data. Meanwhile, the agent can control the detector's position, and human-imposed constraints can be implemented on the reward signal. In this way, we establish a loop in which the agent progressively learns the environment, initially through trial and error (\textit{exploration}), but increasingly taking advantage of the knowledge it has acquired (\textit{exploitation}).

In order to present a tangible picture of the use of this technique in future detectors, while not narrowing our discussion to a particular proposed observatory, we chose sites that are plausible (low seismic activity, sparsely populated, and generally flat), yet not under consideration for any GW-related project.
Site 1 ($42^{\circ} 20' 17.67"$N, $5^{\circ} 43' 52.24"$W) is meant to show the general capabilities of this software, by presenting several small settlements that the model needs to avoid and roads whose crossing would entail an increase in costs. On the other hand, the area is large enough to theoretically fit a detector with an arm length of $\sim 25$ $\textrm{km}$.
Site 2 ($39^{\circ} 11' 18.22"$N, $3^{\circ} 9' 14.12"$W) is smaller, yet considerably less obstructed, as there are only a handful of houses and farms along a single named road. It is also flat enough to fit a locally horizontal $\sim10$ km surface detector. 

Two points are enough to naively characterize an L-shaped interferometer: the vertex position and that of one of its ends. Therefore, the agent has eight possible choices: move each point north, south, east, or west. The agent also has the ability to choose between fine or coarse movement, or staying put, for a total of 17 possible actions per observatory. The model uses Proximal Policy Optimization~\cite{RL_PPO} to learn the appropriate actions to be taken in order to maximize the reward obtained over a fixed set of steps (2000).

Finally, the reward function (the RL way of calling the utility function) codifies the optimization priorities. Following the discussion in the previous section, our main proxy for scientific goals is arm length, with penalties applied for crossing roads, requiring tunnels and/or bridges to be built, and having the ends of the detector too close to the edge of the site\footnote{In our case, the sites are limited by highways and high-speed rail lines, proximity to which is undesirable.}.
A specially severe penalty is applied for configurations considered ``invalid". That is, those interferometers that cross excluded perimeters (settlements, wind farms, quarries, natural reserves, etc) or rivers, or whose arms reach outside the optimization area. These configurations are not prohibited outright, as the model may find better overall configurations by navigating through them, but they carry a more substantial penalty and are not recorded in the ledger for consideration in later stages.

All reward structures introduced so far have only concerned individual interferometers, which on their own have limited capabilities~\cite{Iacovelli_2024_combining_ground_and_surface_detectors, Science_with_ET_2023}. For the network-wide reward, we take the average reward of the two detectors and add an extra term dependent on their relative orientation, $\alpha$. The correct way to think of orientations of a pair of interferometers is through the great circle that unites them. However, recreating an analysis made by~\cite{Science_with_ET_2023} we find that by using the relative north of our sites, we only incur a small deviation of $1.68^{\circ}$. We structure the reward function to favor a relative orientation of $45^{\circ}$. This allows to achieve a near-optimal sky localization capabilities of individual mergers while retaining some sensitivity to stochastic GW backgrounds~\cite{Science_with_ET_2023}.

\section{Results}
\label{sec:results}

\begin{table}[b]
\centering
    \begin{tabular}{|c|c|ccccc|}
    \hline
    \multicolumn{2}{|c|}{} & 2L [km] & Tun. [km]& Supp. [km]& R [Mpc] & $\alpha$ ($^{\circ}$) \\
    \hline
    \multirow{2}{*}{C1}& S1 & 24 & 10.62 & 0 & 696& \multirow{2}{*}{21} \\ 
    & S2 & 15 & 0 & 0 & 449&  \\ \cline{2-7}
    \multirow{2}{*}{C2}& S1 & 30 & 13.52 & 0.18 & 843& \multirow{2}{*}{27} \\ 
    & S2 & 16 & 0 & 0.04 & 477&  \\ \cline{2-7}
    \multirow{2}{*}{C3}& S1 & 44 & 26.33 & 2.48 & 1160& \multirow{2}{*}{52} \\ 
    & S2 & 19 & 8.46 & 0.55 & 564&  \\ \cline{2-7}
    \multirow{2}{*}{Ref.}& Adv. & 8 & 0 & 0 & 195& \multirow{2}{*}{-} \\ 
    & A$^+$ & 8 & 0 & 0 & 347 &  \\ \hline
    \end{tabular}
    \caption{Combined arm length, length of tunneled and supported sections, BNS range and relative orientation of our configurations, Adv. LIGO, and A+.}
    \label{tab:configuration-data}
\end{table}

We trained our model for two million steps and tested it by running a full episode (2000 steps) starting from 12 randomly generated seeds. We saved the 100 most rewarded configurations and displayed (e.g., Fig.~\ref{fig:configuration}) the top 15 of each run. The model followed various different approaches to "solve" the problem at hand. The model jumped back and forth between different and equally valid groups of configurations when optimizing from the same given seed. Upon visual inspection, three families of networks stood out and serve as an illustrative example. 
\begin{itemize}
    \item The first (C1, Fig.~\ref{fig:configuration_1} in Supplemental Material) does not reach the full potential of the sites in terms of length, but avoids tunnels as much as possible and supported sections completely. 
    \item C2 (Fig.~\ref{fig:configuration}) follows a very similar idea as the previous one, but improves both detector lengths in exchange for some supported sections.
    \item Finally, C3 (Fig.~\ref{fig:configuration_3} in Supplemental Material) places scientific goals ahead of engineering challenges by proposing long solutions at the cost of larger underground and supported sections, though it still managed to keep one of the arms of the first site mostly on the surface. Site 1 of this configuration is in line with the optimal arm length considered by~\cite{PhysRevD.99.102004_Detector_optim_for_BNS}.
\end{itemize}

\begin{figure}[t]
      \centering
      \includegraphics[width=0.5\textwidth]{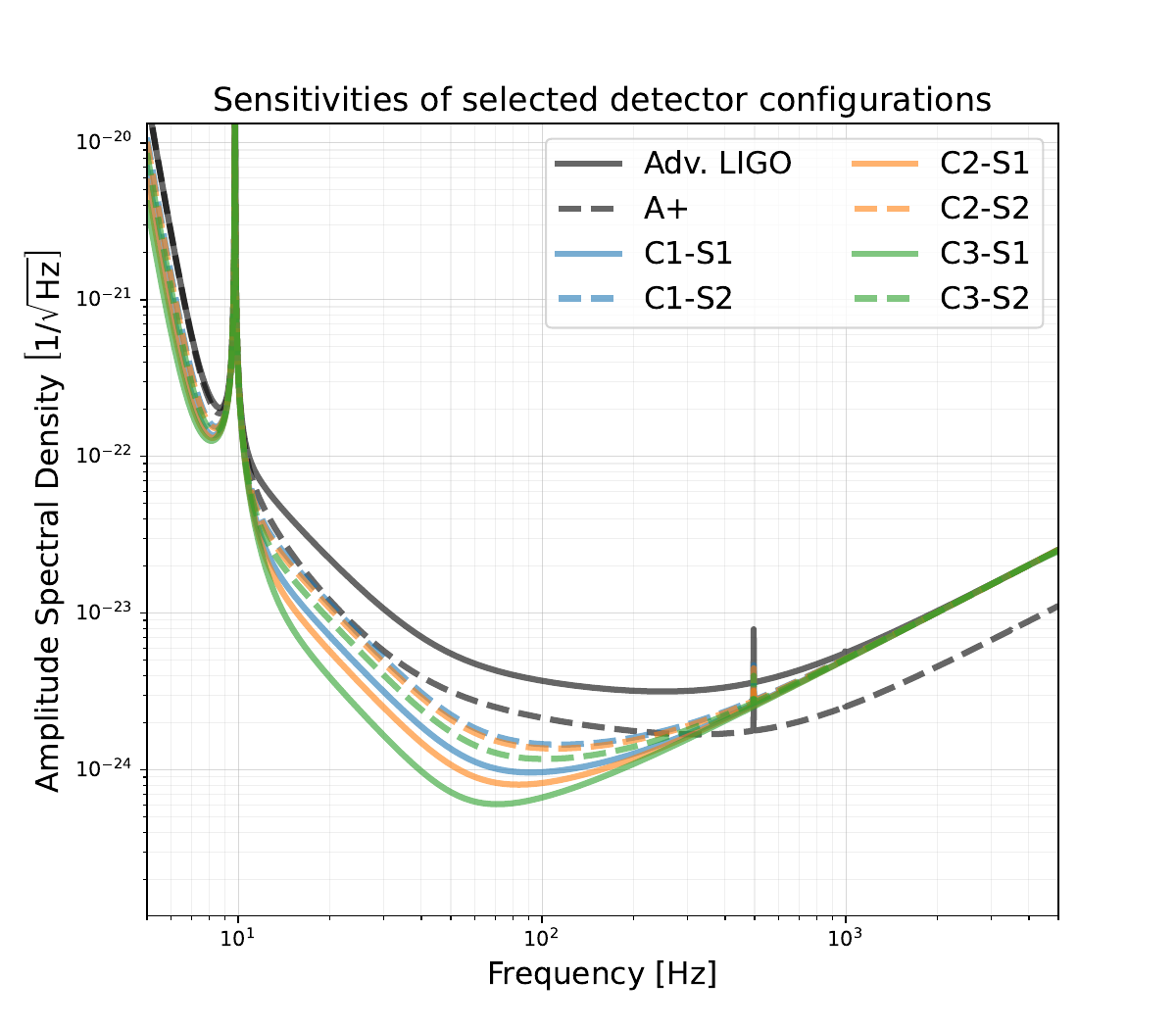}
      \captionof{figure}{Sensitivity curves of the discussed detectors compared to Advanced LIGO~\cite{Advance_LIGO_paper} and the A+ upgrade~\cite{A_plus}.}
    \label{fig:noise_curves}
\end{figure}

Before optimizing the cavity parameters, we slightly modified (by 500 m at most) the configurations, for example, to intersect a road more favorably, or to move further away from populated areas. These micro-adjustments could potentially be performed by a more localized RL simulation with a finer resolution of the terrain, but the importance of humans critically examining ML-suggested designs should not be ignored~\cite{Klimesch2026_Nature_review}. We also rounded the length for convenience, given they are not actual proposals.

Once we have a target length for each of the interferometers, we run the optimization loop described in the methodology until three conditions are met: the length dictated by the RL simulation is achieved, the target cavity g-factor is achieved and the BNS range values no longer improve, all within predetermined tolerances. We set an arbitrary target g-factor of 0.1 and start clipping the gradient of the arm length $50$ m before $L$. 

For this proof-of-concept, we ran \texttt{AutoGrav} starting from the original Advanced LIGO design that comes pre-packaged with \texttt{GWINC}. From it, we only modified the parameters previously discussed  ($L$, $R_{ITM}$, $R_{ETM}$).

We present the resulting sensitivity curves in Fig.~\ref{fig:noise_curves} in comparison to both Advanced LIGO and its planned A+ upgrade. For simplicity, all other independent parameters of the original  design have been left untouched. 
Evidently, future upgrades and detectors will bring technological improvements (and already have~\cite{LIGO_Freq_Dep_Squeezing}), as can be seen in the high-frequency section of the A+ curve. Nevertheless, we show that substantial improvements can be achieved even without them by leveraging RL and DP.

As expected, site 1 is always more sensitive than site 2, and each successive configuration improves over the previous (at a price, of course). While low and high frequencies are bounded by the LIGO design choices, these site-adapted designs outperform Advanced LIGO in the $10$-$800$ Hz range and A+ in the $10$-$300$ Hz range. 

\section{Discussion \& conclusions}
\label{sec:conclusions}

In this work, we presented \texttt{IfoScout}, a new methodology to assist in the concurrent design and placement of a complete network of detectors. To showcase its capabilities, we chose two hypothetical observatory sites and compiled publicly available data about them. Using this data and the LIGO specifications as a starting point, we found a collection of interesting designs that combine scientific ambition with economic viability. 

The fact that a single model trained on a balanced reward function can produce configurations that favor conservative and ambitious proposals could be harnessed by tailoring more specialized reward structures adapted to the needs of each specific project.

While we have presented the problem at hand as a sequential optimization, newtonian and anthropogenic noises are highly sensitive to exact detector position, which, in practice, couples stages 1 and 3 of the pipeline. This non-separability compromises the quality of solutions of sequential optimizations~\cite{MODE:2026xoy}. Merging them into a simultaneous pipeline would require the development of a differentiable clone or surrogate of multiple geometry software suites, a task beyond the scope of this work.

It is also important to note that in this proof-of-concept we have not taken into account various phenomena that a production-ready pipeline must address, one of them being the curvature of the earth for the RL simulation. This is relevant for detectors of this size, but more so for a network of them~\cite{Datrier_2026_CE_sites, Bristol_CE_site_analysis_2026}. Tunnel construction and test mass setup are likely to be the areas most affected by this simplification. The analysis used to gauge the network-wide reward remains valid, but it would be generally preferable for detectors to be further away from each other than discussed here ($\sim2500$ km as per~\cite{Iacovelli2026_baseline_sky_position}.).

The case for optimizing interferometer placement within sites is especially relevant in Europe, where surface detectors of $15\sim20$ km have so far been discarded mostly on the basis of length constraints~\cite{Iacovelli_2024_combining_ground_and_surface_detectors}. However, we show, by means of ML-powered systematic searches of the parameter space, that detectors longer than the currently proposed ones are potentially feasible.  Even though this study has focused only on surface and partially underground\footnote{Solutions whose arms are partially underground are still seismically treated as surface detectors. This represents a worst case scenario and is coherent with the fact that they are, at most, 45 meters underground, almost an order of magnitude closer to the Earth's surface than ET is expected to lie~\cite{ET_site_criteria}.} detectors, this methodology could also be extended for underground detectors, where 2D elevation maps are replaced by 3D geological data and populated areas by aquifers. The necessity of such efforts has already been highlighted~\cite{Science_with_ET_2023}.

After all, the cornerstone of the design philosophy of this pipeline is its flexibility in incorporating the experience and expertise of established collaborations in order to perform realistic reward shaping (e.g. specific safety margins from end stations to highways).  Going forward, we expect to further improve its versatility by including auxiliary facility buildings/caverns, refine the geographical simulation as previously discussed to properly model underground detectors, and start taking into account social factors, crucial to such large projects. These can include factors as diverse as staff housing, commute and amenities, local support for a proposal, and environmental impact, among others. 

For the differentiable simulation, we have introduced a gradient-projection technique that can model the interdependence of parameters at runtime and account for limitations of \texttt{GWINC}. As an example, the g-factor is a proxy for the beam-spot size, which in  \texttt{GWINC} is inversely proportional to the coating thermal noise. Even before compromising the stability of the cavity, coating defects constrain the size a beam-spot can have. This effect is not modeled in  \texttt{GWINC}, but we can reproduce it with a target g-factor hyperparameter set by the mirror coatings team.

Although we have used a looser definition of co-design compared to Ref.~\cite{MODE:2026xoy}, this methodology, in conjunction with the ongoing efforts~\cite{Iacovelli_2022_GWFAST, DupletsaHarms2023_GWFISH, Begnoni_2026_GWJULIA, Santoliquido_2026_Next_gen_NPE, de_Souza_2026_GWDALI_JAX}, also opens the door to a proper study on the systematic hardware-software codesign of GW detectors together with their control and data analysis pipelines.

Our ML-assisted pipeline allows for a faster (see Supplemental Material) and more thorough study of both prospective and familiar detector sites, which is likely to positively impact the deployment timeline and economical cost of GW projects by complementing initiatives for future detectors~\cite{ET_site_criteria, CE_site_criteria, Datrier_2026_CE_sites, Bristol_CE_site_analysis_2026, Science_ET_2026}, as well as upgrades to current ones.  This kind of systematic exploration and evaluation of large portions of the parameter space will also give designers more confidence in their decisions, and therefore in the eventual, human-refined designs. 

\section{Data Availability}
Our work has made use of data, software, and web tools obtained from the Instituto Geográfico Nacional (IGN)~\cite{IGN_MDT02} and Google Earth~\cite{googleearth}.
The code of \texttt{IfoScout}, including a detailed example, is publicly available at~\cite{ifoscout}.

\section{Acknowledgments}

PV is supported by the “Ramón y Cajal” program under Project No. RYC2021-033305-I funded by the MCIN MCIN/AEI/10.13039/501100011033 and by the European Union NextGenerationEU/PRTR, and by the European Innovation Council (EIC) Pathfinder project PHINDER, grant agreement No. 101258353, funded by the European Union. LT acknowledges the Spanish Ministerio de Ciencia, Innovación y Universidades for partial financial support under the projects PID2022-140670NA-I00 and PID2021-125630NB-I00. The authors gratefully acknowledge the computer resources at Artemisa and the technical support provided by the Instituto de Fisica Corpuscular, IFIC (CSIC-UV). Artemisa is co-funded by the European Union through the 2014-2020 ERDF Operative Programme of Comunitat Valenciana, project IDIFEDER/2018/048. 

This work has benefited from activities carried out within the framework of the COST Action MLQC4FC (CA24146), supported by COST (European Cooperation in Science and Technology).

 We also thank J. A. Font and O. G. Freitas for very insightful conversations during the development of this project.

\bibliographystyle{apsrev}
\bibliography{main} 

\clearpage

\begin{figure*}
    \centering
    \subfigure[C1-S1: 254 iterations.]{\includegraphics[width=0.48\textwidth]{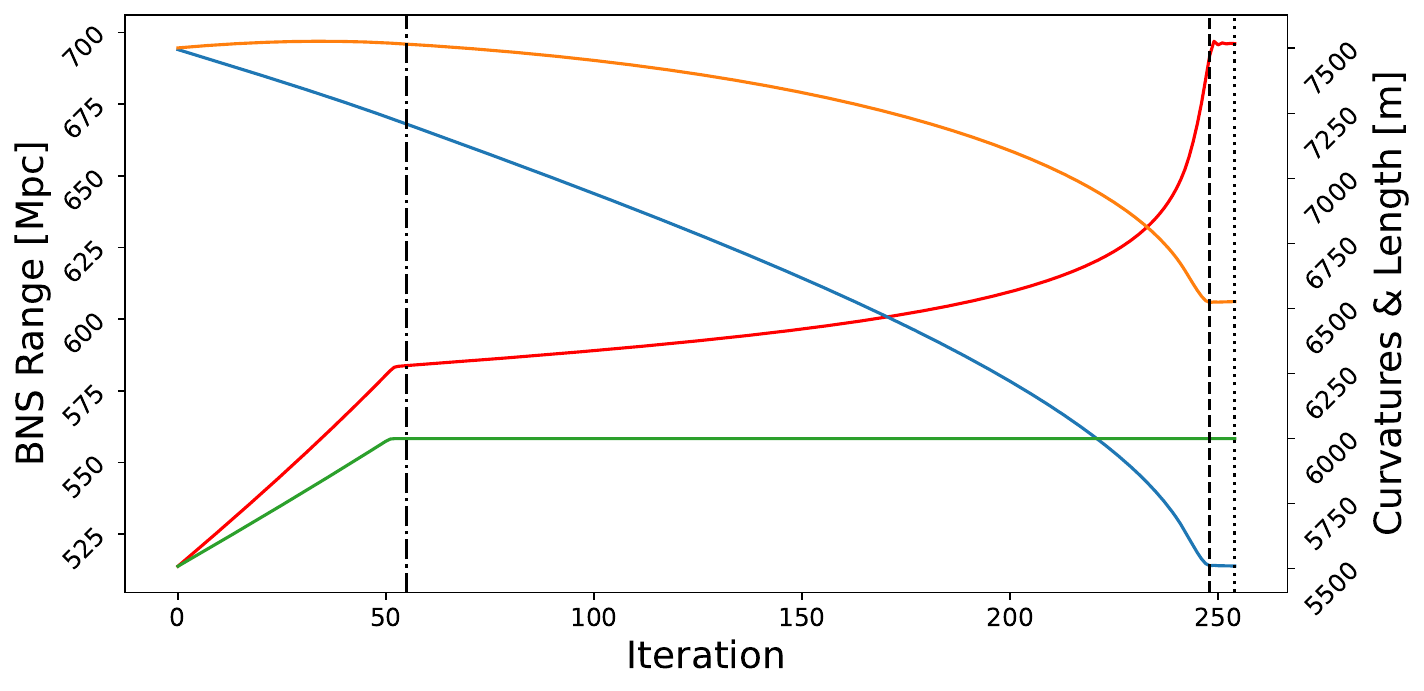}} 
    \subfigure[C1-S2: 90 iterations.]{\includegraphics[width=0.48\textwidth]{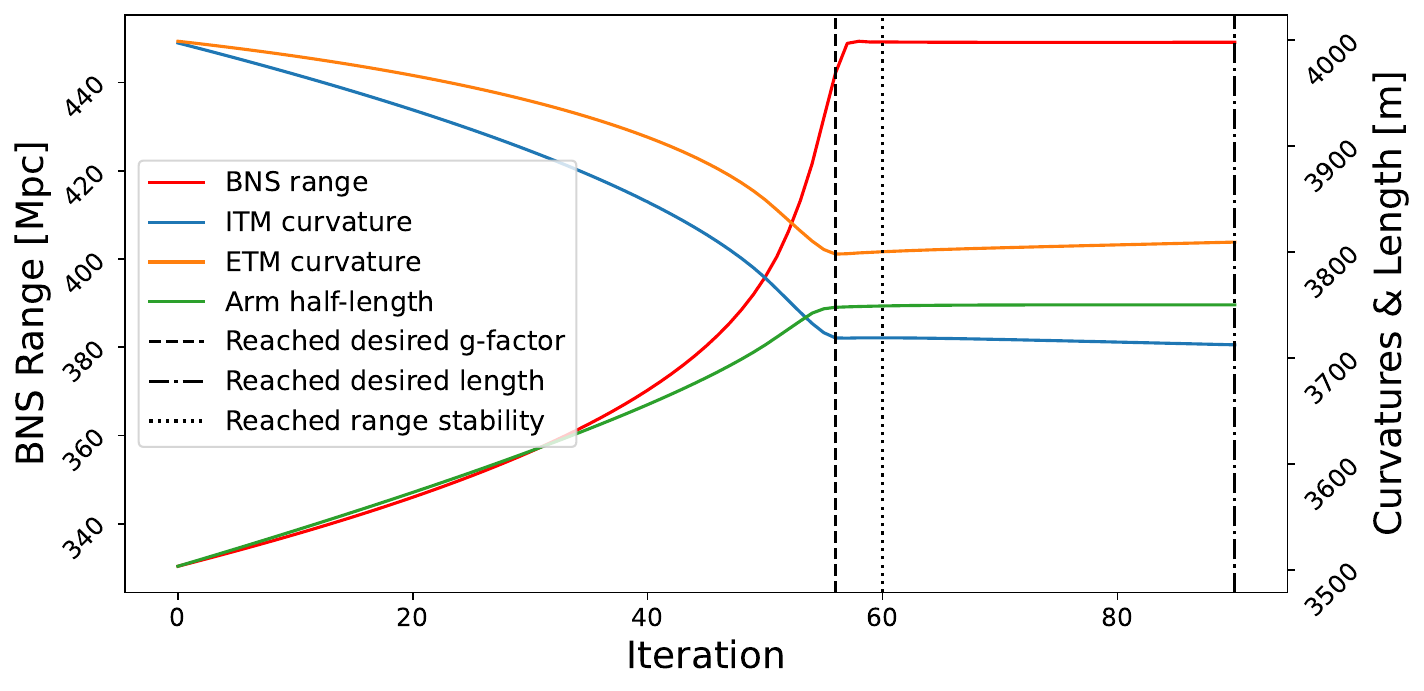}} 
    \subfigure[C2-S1: 155 iterations.]{\includegraphics[width=0.48\textwidth]{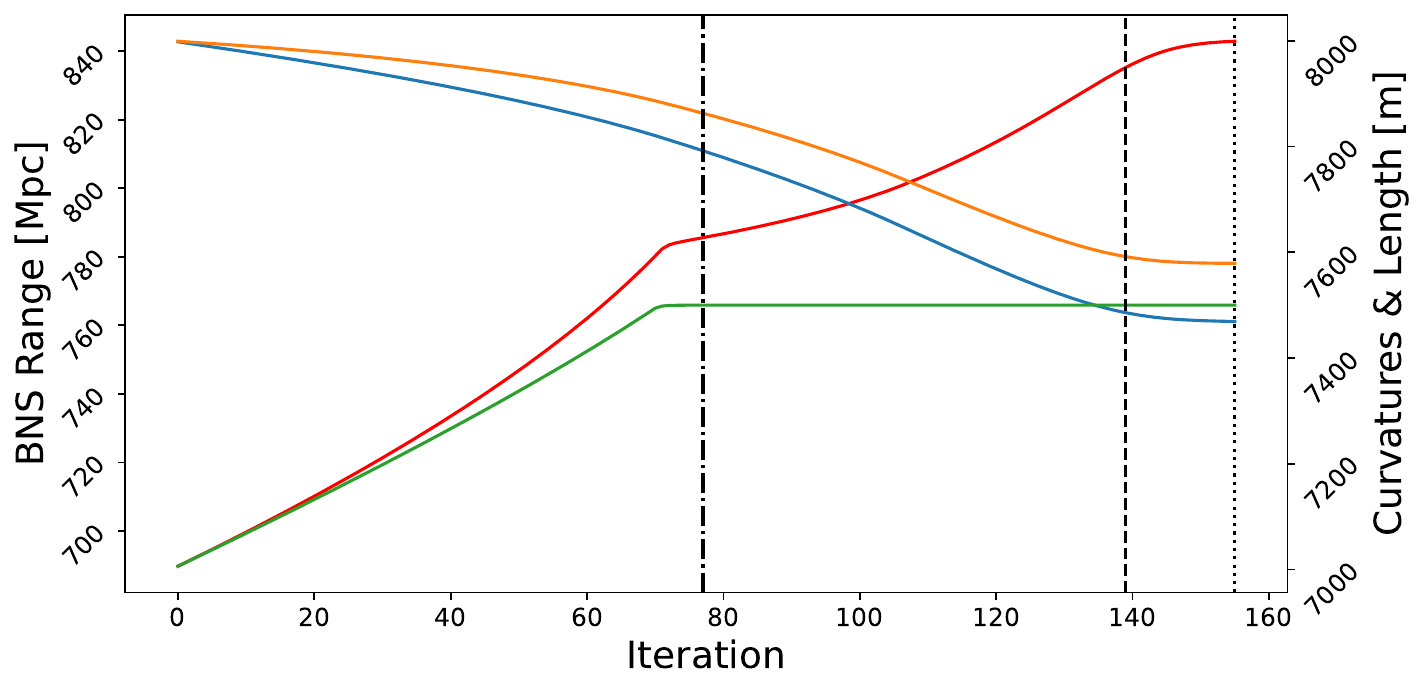}}
    \subfigure[C2-S2: 139 iterations.]{\includegraphics[width=0.48\textwidth]{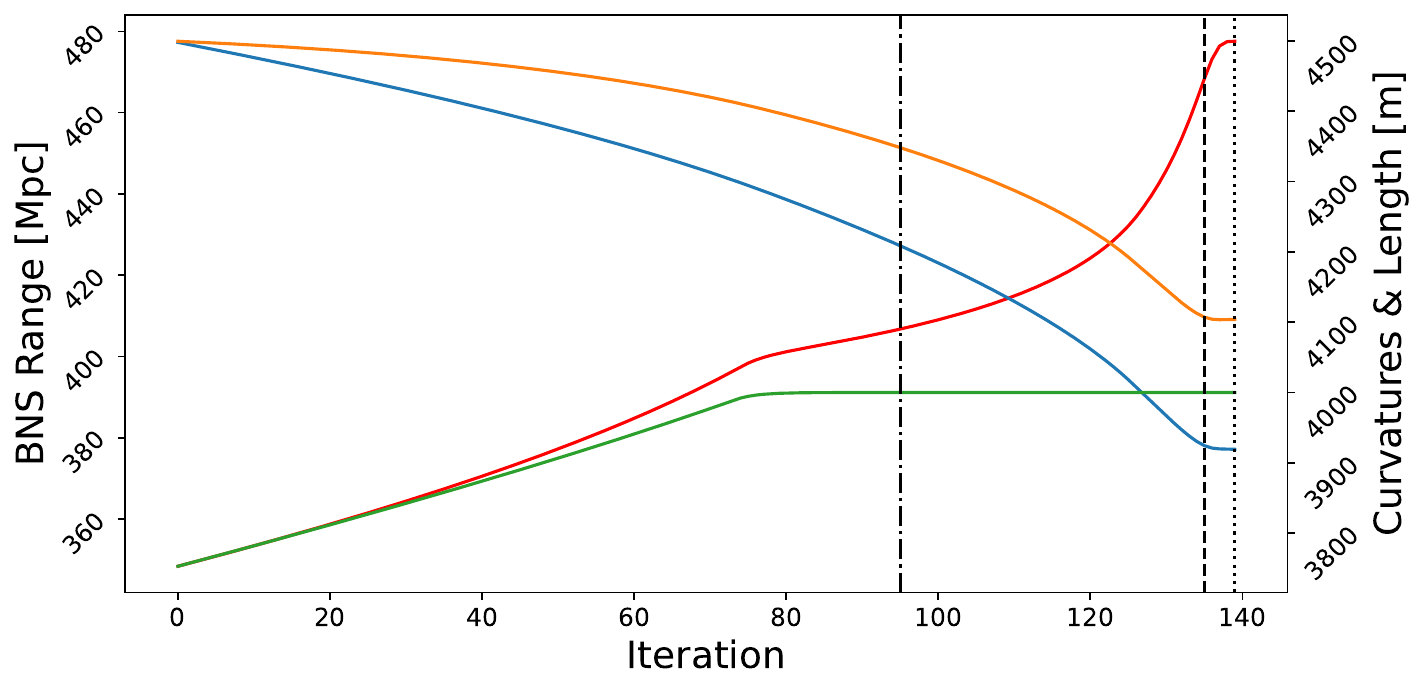}}
    \subfigure[C3-S1: 271 iterations.]{\includegraphics[width=0.48\textwidth]{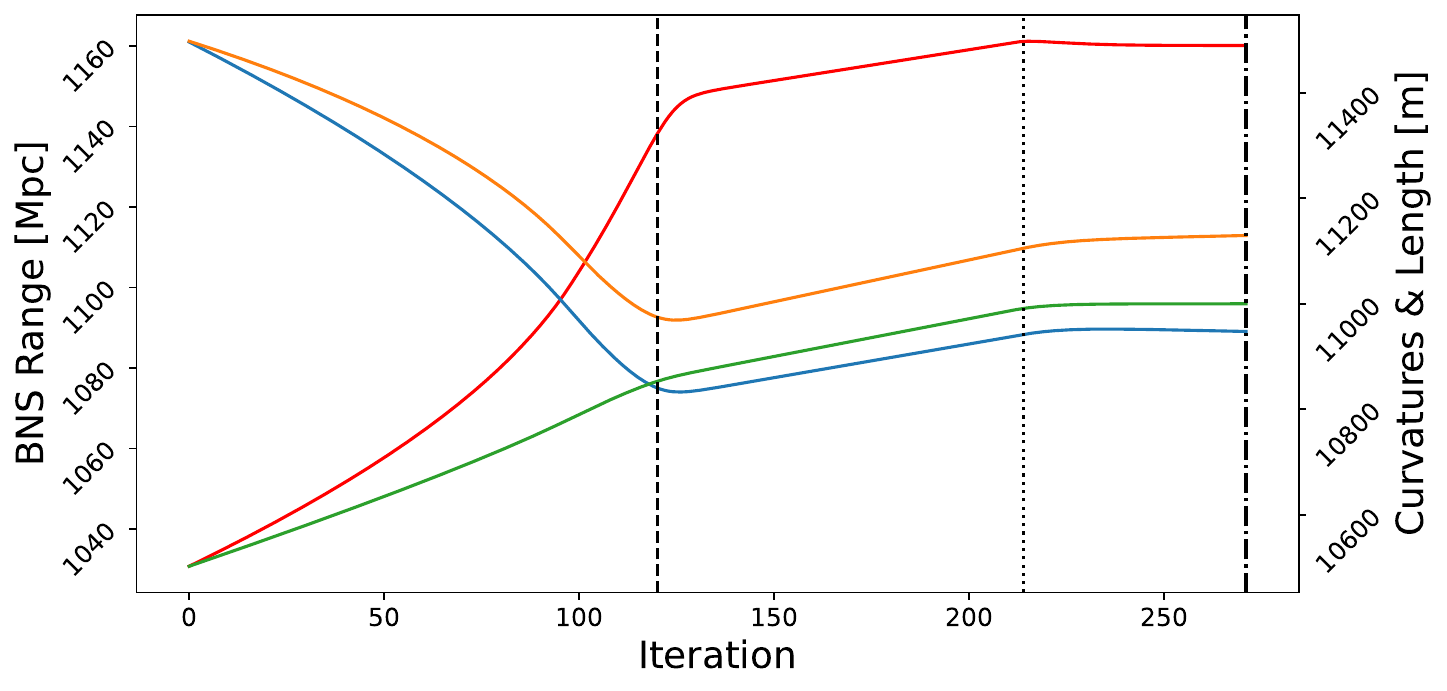}}
    \subfigure[C3-S2: 307 iterations.]{\includegraphics[width=0.48\textwidth]{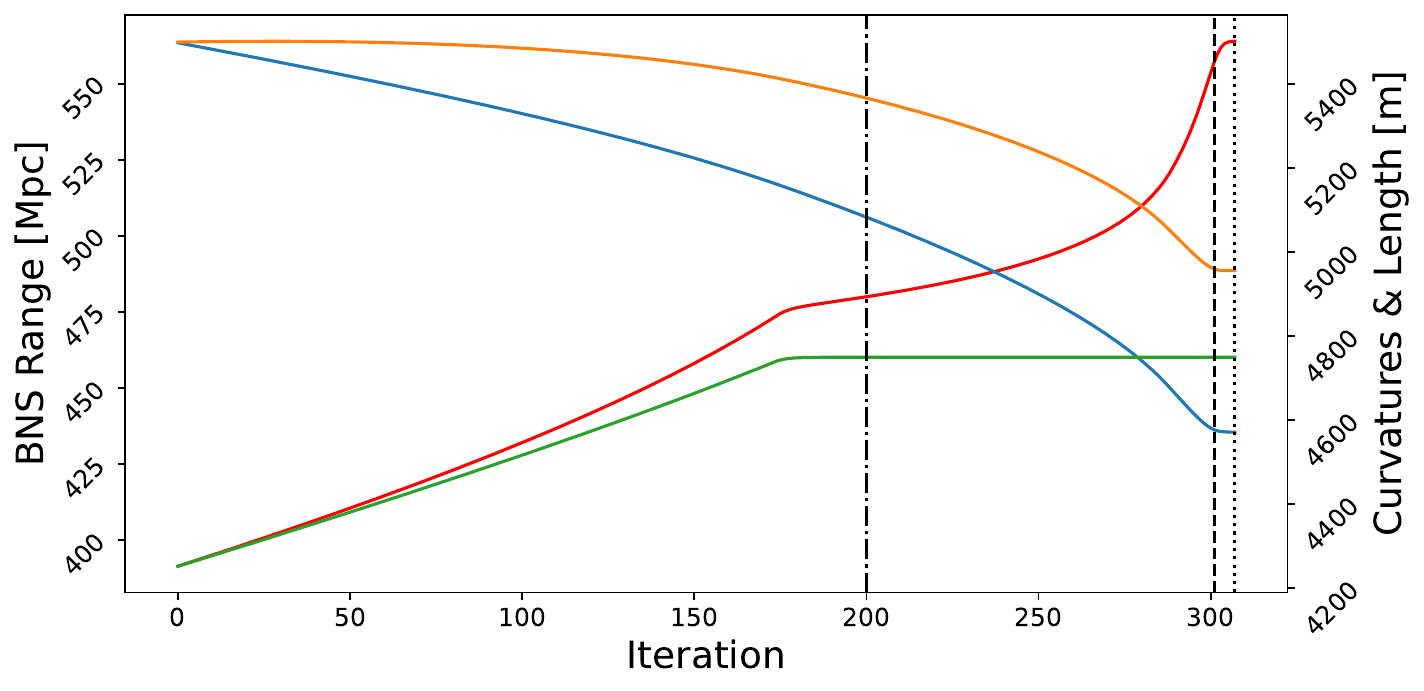}}
    \caption{Optimization diagnostics of the various configurations. All optimizations were run with a constant learning rate of $2\cdot 10^{-5}$. Vertical lines mark the achievement of individual optimization objectives, which need not be reached in a specific order.}
    \label{fig:autograv-diagnostics}
\end{figure*}
\begin{figure*}[t]
      \centering
      \includegraphics[width=\textwidth]{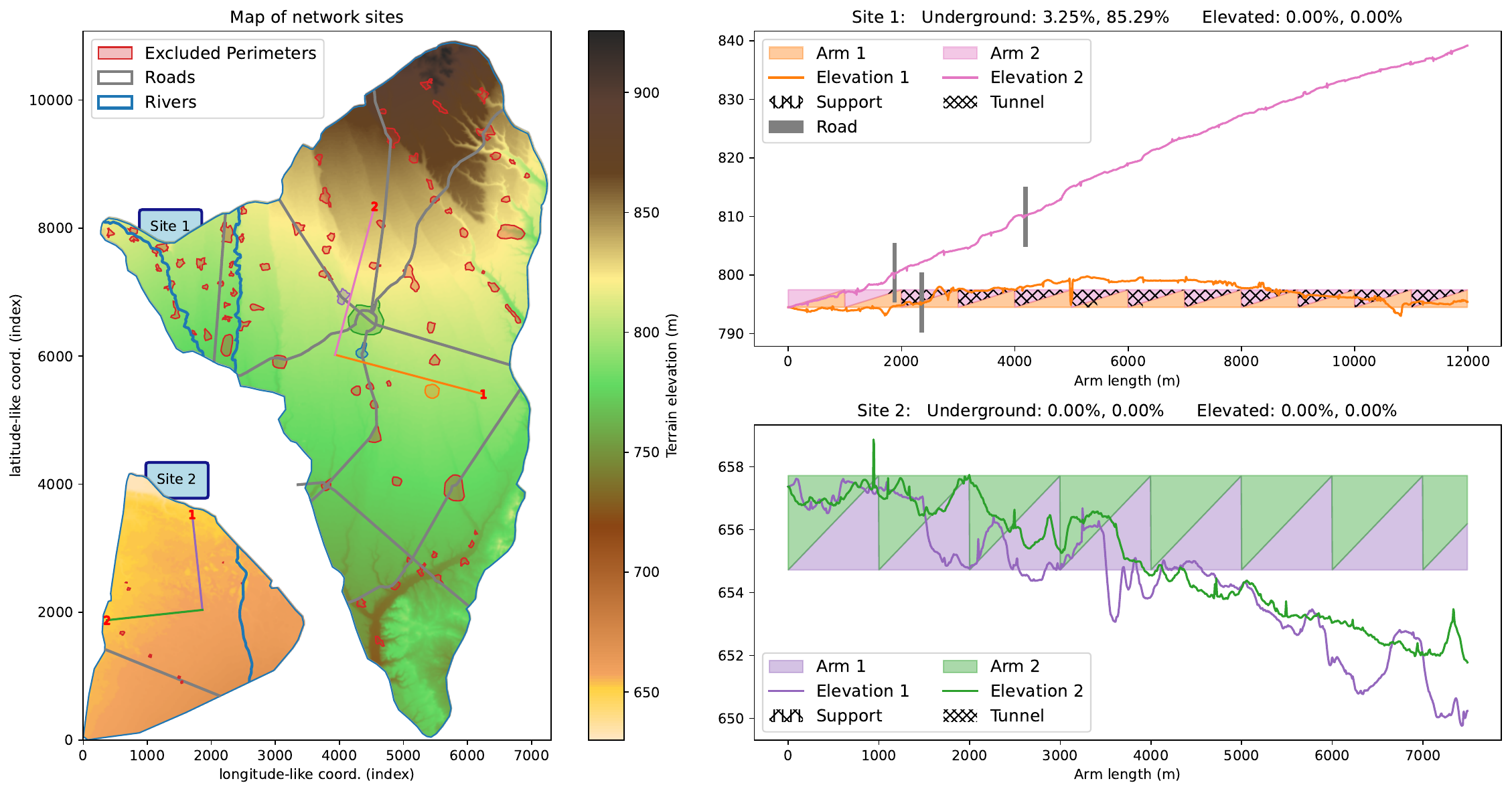}
      \captionof{figure}{Summary plot of Configuration 1. Populated areas discussed on the text appear color-coded.}
    \label{fig:configuration_1}
\end{figure*}

\begin{figure*}[b]
      \centering
      \includegraphics[width=\textwidth]{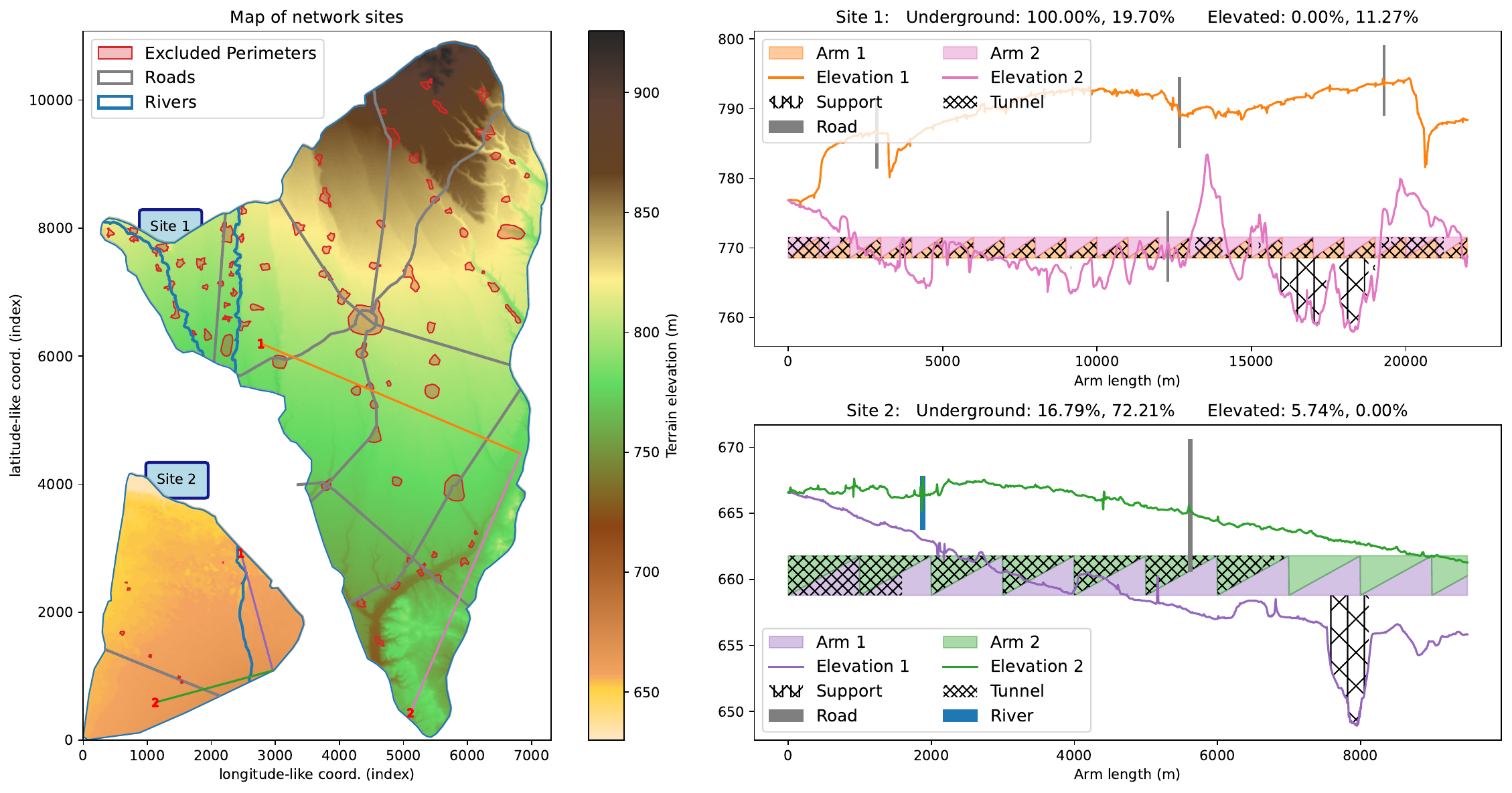}
      \captionof{figure}{Summary plot of Configuration 3. Note the canal near the 2000 meter mark.}
    \label{fig:configuration_3}
\end{figure*}
\section{Supplemental Material}
\subsection{Optimization timescales}
The design phase of projects of this magnitude is often long enough to render design optimization times almost negligible. However, our highly parallelizable approach is particularly fast, with the RL model displayed in this proof-of-concept taking $2\cdot 10^{6}$ steps in less than an hour within 16 environments (64 CPUs) and $\sim 30$ seconds to run a full episode on 12 environments ($2\cdot 10^{3}$ steps, 14-CPU-laptop). Each of the six DP optimizations shown in Fig.~\ref{fig:autograv-diagnostics} took about a minute (14-CPU-laptop). This speed enables a higher degree of experimentation with different utility functions encoding various priorities, as discussed in the main text.

\subsection{Notes on the convergence of \texttt{AutoGrav}}
To declare convergence in the \texttt{AutoGrav} simulation we enforce three requirements through hyper-parameters of the optimization:
\begin{itemize}
    \item Reaching the length set by the RL simulation. We treat length as a variable parameter to allow the optimizer to converge to the optimal solution through its preferred approach, unconstrained by the two-dimensional manifold of actually varying parameters.
    \item Reaching a desired g-factor, ensuring that the optical cavity is as theoretically stable as agreed upon by the design team and the future operators of the instrument. This is a key philosophy of this design approach that will be more vital as detector complexity inevitably increases.
    \item Having a stable BNS range value is a basic expectation of any optimal solution. That is, the absolute difference in range between iterations is smaller than a given tolerance.
\end{itemize}

All six optimizations start from relatively similar positions: Length is set back $500\sim 1000$ m from the target, and both curvature radii are initially equal and marginally above $L/2$ to ensure initial compliance with Eq. \ref{eq:g-factor}.  For every iteration, the gradient is obtained and multiplied by a constant "learning rate" (for its analog in machine learning) of $2\cdot 10^{-5}$ before any gradient projection is applied.

As seen in Fig.~\ref{fig:autograv-diagnostics}, some interferometers reached the desired g-factor before the desired length. However, most reached the length first, then adapted the mirrors for the specified cavity stability, and then settled on a stable set of parameters. Despite these different morphologies on the optimization process, the solutions found also present many shared characteristics. Specifically, we draw attention to the fact that all configurations fall into a pattern $R_{ITM} < L < R_{ETM}$, common among operating detectors~\cite{Advance_LIGO_paper, Advance_Virgo_paper}, in spite of the shared initial start of the curvature radii and their symmetrical constraints being deliberatively agnostic to this fact.

\subsection{The structure of the utility function}

The heart of the RL simulation is its reward/utility function, as it indicates the priorities the agent must learn. However, it is also highly customizable and dependent on the priorities of each project, so the nature of its inclusion in Eq.~\ref{eq:utility} is to provide context on the inner workings of this particular demonstration, as opposed to \texttt{GWymnasium} as a whole.
\begin{equation}\label{eq:utility}
    U(L, L_{-1}; \vec{d}, \vec{f}) = \alpha L + \tilde{\alpha} (L-L_{-1}) - \sum_{\textrm{pen.}} \beta_iP_i(\vec{d}, \vec{f})
\end{equation}
The reward has two positive terms, dependent only on detector length, $L$, and that of the previous step, $L_{-1}$. The first prioritizes long detector arms as a proxy for sensitivity, while the second one encourages growth from one configuration to the next, which made the model achieve remarkable configurations faster.

The rest of the dependencies on the environment are codified in the form of penalties. We gather two types of information from the environment: data, $\vec{d}$, such as the distance of the vertices and edges of the interferometers to the edge of the site, or the fraction of the arms that would require tunneling or support. As alluded to in the main text, we use the latter as a proxy for the economical construction cost for this example, penalizing it. An actual project could potentially remain at a fixed budget.
The other type is flags, $\vec{f}$. Discrete pieces of information that relay to the model whether it has crossed a particular region, a given road any number of times, or if the current configuration is valid. These encode various social and political priorities such as avoiding populated areas or penalizing road disruption.

Finally, the utility function of each member of the network is summed and the orientation-dependent reward, $O_{ij}$, discussed in the main text, is added in Eq.~\ref{eq:network_util}. Since this simple demonstration only has two interferometers, there is a single orientation reward. For information on specific weights ($\alpha$,$\tilde{\alpha}$, $\beta_i$ and $\gamma$) see the example in Ref.~\cite{ifoscout}.
\begin{equation}\label{eq:network_util}
    U_{net}(U_i, \alpha_{ij}) = \frac{1}{N}\sum_i^NU_i + \sum_{i<j}\gamma O_{ij}(\alpha_{ij})
\end{equation}

\subsection{Additional details on the obtained configurations}

With this being a fictitious observatory, the exact configurations themselves are relatively unimportant. However, the solutions that the optimizer found in order to fit the detectors in their respective sites are worth exploring further.
\begin{itemize}
    \item C1 (Fig.~\ref{fig:configuration_1}) managed to be economical by avoiding extensive crossing structures. The program found a stretch of land on site 1 that was, for over 96\% of its length, within the 5 meter vertical margin that we had stipulated as the limit for light earthwork solutions. Fitting the first arm of detector along that path required placing the other northbound, against the incline. Even so, the solution is threaded between the populations of Laguna Dalga (blue, population 608) and Pobladura de Pelayo García (orange, pop. 364) for arm 1 and Santa María (green, pop. 2994) and Urdiales del Páramo (purple, pop. 445) for arm 2 (always within the configured margins) with a vertex facility on the surface and only around 10 km of main tunnel for a total arm length of 24 km.

Site 2 has, as established before, a significantly flatter profile than site 1. As expected, the model had no trouble finding a configuration with no crossing structures at all. Only some light soil filling would be necessary over the last 2.5 km of either arm.
    \item The balanced solution (C2, Fig.~\ref{fig:configuration}) slightly enlarges the concept of C1 in both sites. Arm 2 is no longer as tightly threaded between populations, and arm 1 now passes south of Pobladura instead of north. This produces a larger, more sensitive detector, but at a higher cost. Both sites would have required a small supported section, although they may have been low and small enough for the design team to opt for light earthworks.
    \item The more scientifically ambitious (C3, Fig.~\ref{fig:configuration_3}) is substantially different from the other two in its approach. It manages larger detectors by passing over small valleys and tunneling under hills, resulting in profiles with intermittent, yet considerable, supported and tunneled sections, such as site 1, arm 2 (pink). Still, this arm remains mostly on the surface, in spite of its impressive 22 km. Arm 2 of the other site appears to have an undesirable complication: $2$ km from the vertex, it passes under ($\sim$ 8 meters below) a river, which in most circumstances would disqualify this configuration. However, upon visual inspection of the site, we learned that this geographic feature is actually a canal only a couple meters deep that remains empty or with little flow rate throughout the year. We therefore lowered the exclusion threshold for rivers in this site to 5 meters below its surface, as visible in the length of the vertical line representing the intersection of the profile and the picture in Fig.~\ref{fig:configuration_3}. In the future, we would like to extend that level of control to individual features, as opposed to the current class-wide system.
\end{itemize}

\end{document}